\documentclass[pra,aps,twocolumn]{revtex4-1}
\usepackage{amsmath}
\usepackage{amssymb}
\usepackage{graphicx}
\usepackage{bbold}

\newcommand{\beqa}{\begin{eqnarray}}
\newcommand{\eeqa}{\end{eqnarray}}

\begin{document}
\title{PT symmetric lattices with a local degree of freedom} 
\author{Harsha Vemuri}
\author{Yogesh N. Joglekar}
%\email{yojoglek@iupui.edu}
\affiliation{Department of Physics, 
Indiana University Purdue University Indianapolis (IUPUI), 
Indianapolis, Indiana 46202, USA}
\date{\today}
\begin{abstract}
Recently, open systems with balanced, spatially separated loss and gain have been realized and studied using non-Hermitian Hamiltonians that are invariant under the combined parity and time-reversal ($\mathcal{PT}$) operations. Here, we model and investigate the effects of a local, two-state, quantum  degree of freedom, called a pseudospin, on a one-dimensional tight-binding lattice with position-dependent tunneling amplitudes and a single pair of non-Hermitian, $\mathcal{PT}$-symmetric impurities. We show that if the resulting Hamiltonian is invariant under exchange of two pseudospin labels, the system can be decomposed into two uncoupled systems with tunable threshold for $\mathcal{PT}$ symmetry breaking. We discuss implications of our results to systems with specific tunneling profiles, and open or periodic boundary conditions. 
\end{abstract}
\maketitle
%---------------------------------------------------------------------------------%

\noindent {\it Introduction:} A non-Hermitian Hamiltonian $H\neq H^{\dagger}$ that is invariant under the combined parity and time-reversal ($\mathcal{PT}$) operations is called $\mathcal{PT}$ symmetric. Since the groundbreaking discovery of such Hamiltonians in continuum models fifteen years ago~\cite{bender1}, significant research has been carried out to identify and characterize the properties of $\mathcal{PT}$-symmetric Hamiltonians that, typically, are decomposed into a Hermitian kinetic term and a non-Hermitian, $\mathcal{PT}$-symmetric potential term, $V(x)=V^*(-x)\neq V^\dagger(x)$~\cite{bender2,bender3}. The spectrum $\epsilon_\lambda$ of such a non-Hermitian Hamiltonian is purely real over a region of the parameter space, called the ${\mathcal PT}$-symmetric phase; in this region, its (non-orthonormal) eigenvectors $|\phi_\lambda\rangle$ are simultaneous eigenfunctions of the $\mathcal{ PT}$ operation. For parameters outside the $\mathcal{PT}$-symmetric phase, the eigenvalues of the Hamiltonian occur in complex conjugate pairs, and due to the anti-linear nature of the time-reversal operator $\mathcal{T}$, the corresponding eigenfunctions are not simultaneous eigenfunctions of the $\mathcal{PT}$ operation. This emergence of complex eigenvalues is called $\mathcal{PT}$-symmetry breaking. $\mathcal{PT}$-symmetric Hamiltonians are ideally suited to model non-equilibrium phenomena that transition from a quasi steady-state behavior ($\mathcal{PT}$-symmetric phase) to loss of reciprocity (broken $\mathcal{PT}$-symmetry)~\cite{kottos1,kottos2}. 

In the past three years, experiments on coupled optical waveguides~\cite{expt1,expt2,expt3,expt4}, coupled electrical circuits~\cite{rc}, and coupled pendulums~\cite{pendulum} have shown that instead of being a mathematical curiosity, ${\mathcal PT}$-symmetric Hamiltonians represent open (quantum) systems with spatially separated, balanced, loss and gain. The discrete nature of these systems has also sparked new interest in the properties of $\mathcal{PT}$-symmetric tight-binding lattice models with different topologies~\cite{znojil1,znojil2}; such lattice models are most readily realized in evanescently coupled optical waveguides~\cite{yariv,review}. Recent theoretical work has led to the identification of robust and fragile $\mathcal{PT}$-symmetric phases in a lattice with open boundary conditions~\cite{bendix,song,mark,localpt}, tunable $\mathcal{PT}$-symmetric threshold in a lattice with periodic boundary conditions~\cite{derekring}, and substantially strengthened $\mathcal{PT}$-symmetric phase in finite lattices with position-dependent tunneling profile~\cite{avadh,evenodd}. All of this work is, however, restricted to systems in one spatial dimension where the parity operation is defined as $\mathcal{P}: x\rightarrow -x$ in the continuum case (with a suitably defined origin) and $\mathcal{P}: k\rightarrow\bar{k}=N+1-k$ in a lattice with $N$ sites. In particular, the properties of $\mathcal{PT}$-symmetric Hamiltonians in two (or higher) dimensions have been barely explored~\cite{pt2d}. 

In this paper, we investigate $\mathcal{PT}$-symmetric lattices with a local, two-state, quantum degree of freedom labeled by a pseudospin $\sigma=\pm 1$. We present a class of models that can be mapped onto one-dimensional lattice models that have been investigated in the past, and thus are solvable in a straightforward manner. Such degree of freedom can represent, for example, two orthogonal polarizations of a mode in a single elliptical waveguide~\cite{elliptical} in an array of coupled elliptical waveguides. Thus, although we use the term ``pseudospin'' to denote this degree of freedom, we emphasize that its time-reversal properties are unspecified. By using physically motivated $\mathcal{PT}$-symmetric potentials (at only two sites) and tunneling amplitude profiles, we show that the local degree of freedom leads to a robust, tunable $\mathcal{PT}$-symmetric phase. Although our results are applicable to general $\mathcal{PT}$-symmetric systems, in the following, we use a language that is applicable to coupled optical waveguides~\cite{review}. 

%---------------------------------------------------------------------------------%

\begin{figure}[t!]
\begin{center}
\includegraphics[angle=0,width=\columnwidth]{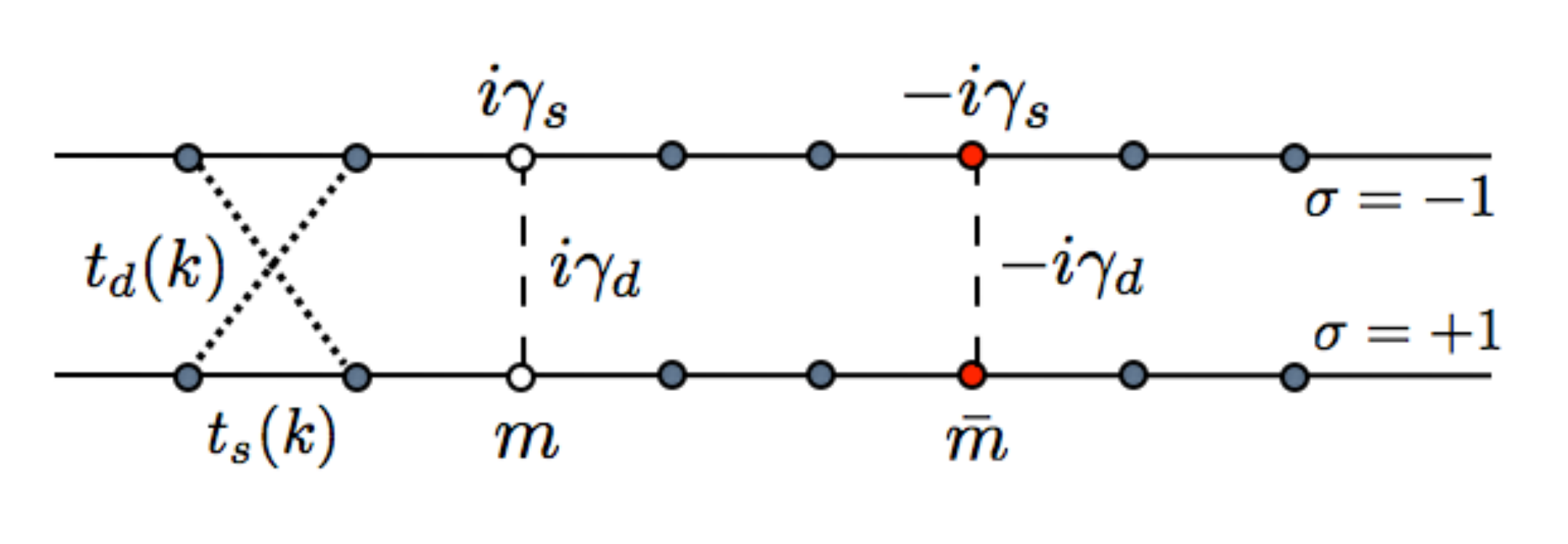}
\caption{(color online) Tight binding lattice with a local degree of freedom represented by pseudospin $\sigma=\pm 1$. The tunneling amplitudes $t_s(k)$ and $t_d(k)$ denote processes that preserve or change the pseudospin respectively, while the tunneling from site $k$ to $k+1$. Similarly, $\pm i\gamma_s$ and $\pm i\gamma_d$ are on-site, non-Hermitian potentials that represent gain (white site) or loss (red site). This system is $\mathcal{PT}$-symmetric irrespective of the time-reversal properties of the pseudospin.}
\label{fig:schem}
\end{center}
\vspace{-5mm}
\end{figure}

\noindent{\it Tight-binding Model:} We consider a lattice of $N$ coupled waveguides with open boundary conditions, described by the following tight-binding Hamiltonian, 
\begin{equation}
\label{eq:tb}
H_0=-\sum_{i=1}^{N-1} \left[ a^\dagger_i T(i) a_{i+1} + a^\dagger_{i+1} T^\dagger(i) a_i\right]. 
\end{equation}
Here $a^{\dagger}_k=(a^\dagger_{k,+},a^\dagger_{k,-})$ are the creation operators for the two modes $|k,+\rangle$ and $|k,-\rangle$ at site $k$ respectively. $\left[T(k)\right]=t_s(k)\mathbb{1} +t_d(k)\tau_x$ is the 2$\times$2 tunneling matrix that couples the two modes at site $k$ to the two modes at site $k+1$, $\tau_x$ is the $x$-Pauli matrix, and $t_s\geq 0$ ($t_d\geq 0$) denote the tunneling amplitude for processes that preserve (flip) the local degree of freedom (Fig.~\ref{fig:schem}). We choose a parity-symmetric, real tunneling function $t(k)=t(N-k)$ to ensure that $H_0$ commutes with the combined $\mathcal{PT}$ operator~\cite{avadh}. Note that we have chosen tunneling matrix so that Eq.(\ref{eq:tb}) is invariant under the exchange of pseudospin labels $\sigma\leftrightarrow-\sigma$. For a pair of balanced loss or gain impurities at mirror-symmetric locations, the potential is given by 
\begin{equation}
\label{eq:v}
V= a^\dagger_m i\Gamma a_m - a^\dagger_{\bar{m}} i\Gamma a_{\bar{m}}
\end{equation}
where $\bar{m}=N+1-m$, $i\left[\Gamma\right]=i\gamma_s\mathbb{1}+i\gamma_d\tau_x$ denotes the non-Hermitian gain matrix at site $m$, and $0\leq \gamma_d\leq \gamma_s$ denote the gain amplitudes for mode preserving and mode exchanging processes. The potential $V$ is also invariant under the exchange of pseudospin labels, and is $\mathcal{PT}$ symmetric irrespective of the time-reversal properties of the pseudospin. 
% Schematic lattice model.

The eigenvalue difference equation obeyed by a two-component eigenfunction $\Psi(k)=\left(f_k,g_k\right)^{T}$ with energy $\epsilon$ is given by
\begin{eqnarray}
\label{eq:eigen}
-T(k-1)\Psi(k-1)-T(k)\Psi(k+1) & + & \nonumber \\
(\delta_{k,m}-\delta_{k,\bar{m}})i\Gamma\Psi(k)& = &\epsilon\Psi(k).
\end{eqnarray}
where $k=1,\cdots,N$.  We note that open boundary conditions are implemented by assigning $T(0)=0=T(N)$ whereas periodic boundary conditions imply 
$T(0)=T(N)\neq 0$. Using the symmetric and antisymmetric basis that {\it diagonalizes the tunneling matrix $T(k)$ at every site}, it is straightforward to obtain the following decoupled equations, 
\begin{eqnarray}
\label{eq:plus}
-( t^{S}_k f^S_{k+1} + t^S_{k-1}f^S_{k-1} )+i\gamma^S f^S_k(\delta_{k,m}-\delta_{k,\bar{m}}) = & \epsilon f^S_k, &\,\\
\label{eq:minus}
-( t^{A}_k f^A_{k+1} + t^A_{k-1}f^A_{k-1} )+i\gamma^A f^A_k(\delta_{k,m}-\delta_{k,\bar{m}}) = & \epsilon f^A_k. &\,
\end{eqnarray}
Here $t^{S(A)}_k=\left[t_s(k)\pm t_d(k)\right]$ are the symmetric and antisymmetric combinations of the tunneling rates, $\gamma^{S(A)}=(\gamma_s\pm\gamma_d$), and $f^{S(A)}_k=(f_k\pm g_k)$ are the eigenfunction components in the symmetric-antisymmetric basis.

Eqs.(\ref{eq:plus})-(\ref{eq:minus}) show that the $\mathcal{PT}$-symmetric Hamiltonian is {\it a direct sum of Hamiltonians for two lattices with no internal structure}: $H=H_0+V=H_S\oplus H_A$ where $H_S$ is the $\mathcal{PT}$-symmetric Hamiltonian with tunneling profile $t^S_k$ and a pair of non-Hermitian impurities at mirror-symmetric locations $(m,\bar{m})$ with strength $\gamma^S$, and $H_A$ is obtained correspondingly. We emphasize that this decomposition into uncoupled problems is valid for arbitrary, position-dependent tunneling profiles $t_s(k)$, mode-mixing amplitudes $t_d(k)$, open or periodic boundary conditions, and arbitrary loss or gain strengths, as long as the underlying Hamiltonian is invariant under the exchange pseudospin indices. 

%---------------------------------------------------------------------------------%

\noindent{\it Specific Cases and Numerical Results:} When there is no mixing between the two pseudospin states, $t_d=0=\gamma_d$, the problem is trivial. In general, the $\mathcal{PT}$-symmetric threshold for $H_0$ is equal to the smaller of the corresponding thresholds for $H_S$ and $H_A$. 

% Schematic ring model.
\begin{figure}[t!]
\begin{center}
\includegraphics[angle=0,width=0.80\columnwidth]{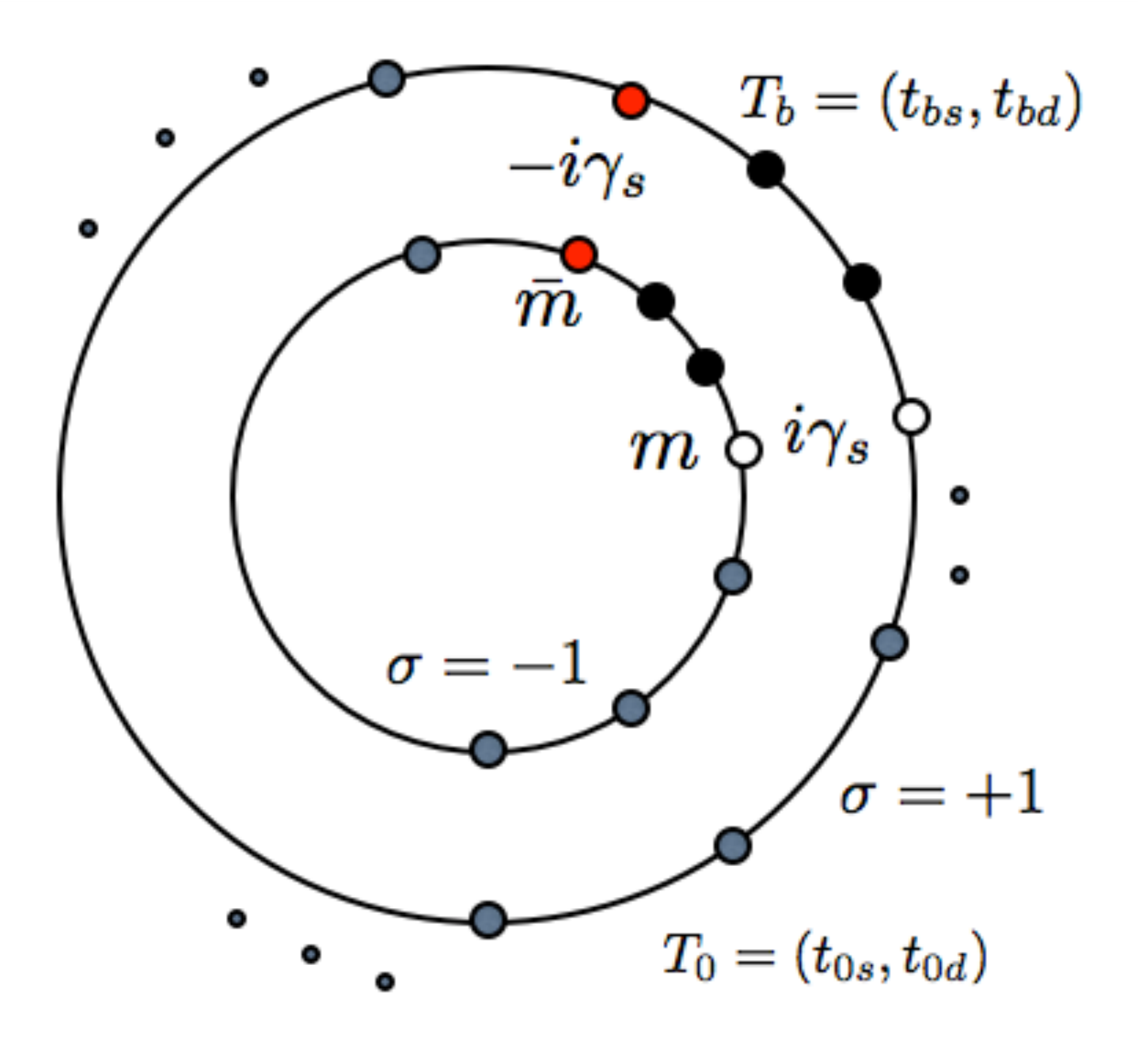}
\caption{(color online) Tight binding ring with a local degree of freedom represented by pseudospin $\sigma=\pm 1$ with a pair of $\mathcal{PT}$-symmetric impurities $\pm i\gamma$ located at arbitrary sites $(m,\bar{m})$. The tunneling matrix $T$ is constant, but different, along the two paths that go from the gain site $m$ (white site) to the loss site $\bar{m}$ (red site). This system will map onto two, uncoupled rings with constant tunnelings $t_0^{S(A)}=(t_{0s}\pm t_{0d})$ and $t^{S(A)}_b=(t_{bs}\pm t_{bd})$ along the two, outer and in-between, paths from site $m$ to site $\bar{m}$.}
\label{fig:shortring}
\end{center}
\vspace{-5mm}
\end{figure}

When $\gamma_d=0$, the loss (or gain) potential couples maximally to the pseudospin eigenmodes $\sigma=\pm 1$, and {\it not to a linear combination of them}. In this case, if the tunneling is constant, the $\mathcal{PT}$-symmetric phase diagram $\gamma_{PT}(m)$ is given by a U-shaped curve, obtained in Ref.~\cite{mark}, with the maximum value $\gamma_{PT}=(t_s-t_d)$. For parity-symmetric, non-constant tunneling profiles, the appreciably strong $\mathcal{PT}$- symmetric threshold, obtained in Ref.~\cite{evenodd}, is now selectively suppressed by increasing the mode-mixing tunneling amplitude $t_d(k)$. For a lattice with periodic boundary conditions, we consider the model with tunneling matrices $T_0=t_{0s}\mathbb{1}+t_{0d}\tau_x$ and  $T_b=t_{bs}\mathbb{1}+t_{bd}\tau_x$ that are constant along each of the two paths that connect the gain site to the loss site (Fig.~\ref{fig:shortring}). It then follows that the $\mathcal{PT}$-symmetric threshold is independent of the distance between the loss and gain sites, as discussed in Ref.~\cite{derekring}, and is given by the smaller of the two combinations, $(t^S_{0}-t^S_{b})$ and $(t^A_{0}-t^A_{b})$. Thus, a $\mathcal{PT}$-symmetric ring with a local degree of freedom offers significant threshold tunability independent of the distance between the loss and gain impurities. 

% Sample numerical results
\begin{figure*}[t!]
\begin{center}
\includegraphics[angle=0,width=2\columnwidth]{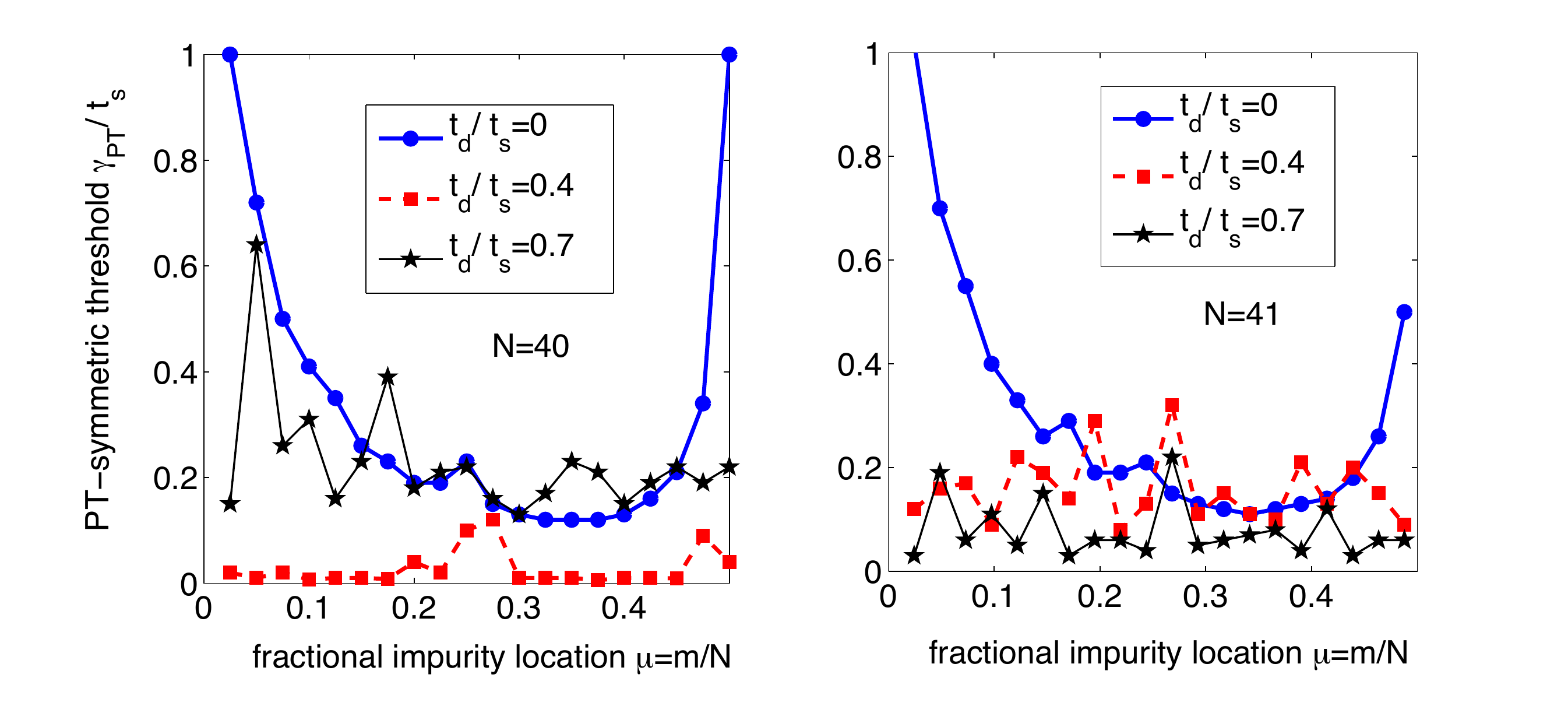}
\caption{(color online) Typical $\mathcal{PT}$-symmetric phase diagram for even ($N=40$, left panel) and odd ($N=41$ right panel) lattices with a local degree of freedom $\sigma$. 
The vertical axis shows the $\mathcal{PT}$-threshold $\gamma_{PT}(\mu)$, measured in units of $t_s$ and the horizontal axis shows the fractional position $\mu=m/N$ of the 
impurity. The tunneling is given by $t_s\geq t_d\geq 0$, and impurity at site $m$ acts as a gain for $\sigma=+1$ and a loss for $\sigma=-1$.  When the mode-mixing $t_d=0$ (blue circles), decoupled pseudospins lead to earlier results~\cite{mark}. As $t_d$ increases (red squares and black stars), the $\mathcal{PT}$-symmetric threshold $\gamma_{PT}(\mu)$ generally decreases.}
\label{fig:phase}
\end{center}
\vspace{-5mm}
\end{figure*}
When $\gamma_d\neq 0$, the analysis carried out here predicts bounds on the gain matrix, given by $(\gamma_s+\gamma_d)\leq (t_s+t_d)$ and $(\gamma_s-\gamma_d)\leq (t_s-t_d)$; however, these bounds do not determine the individual thresholds for $\gamma_s$ and $\gamma_d$. In the extreme case of $\gamma_s=\gamma_d$ (meaning the gain potential only couples to the symmetric combination), we find that $\gamma^A=0$, $H_A$ is a purely {\it Hermitian} Hamiltonian and therefore, the $\mathcal{PT}$-symmetric threshold is solely determined by the Hamiltonian $H_S$. Note that, in general, a direct-sum decomposition of the Hamiltonian $H$ is possible {\it if and only if} the tunneling matrix $T(k)$ {\it at every site $k$} and the non-Hermitian potential matrix $i\Gamma$ can be simultaneously diagonalized. 

Finally we consider the case where the full Hamiltonian $H$ cannot be decomposed into two non-interacting pieces. Generically, for an open lattice or a ring with constant tunneling matrix $T$ and a single pair of gain/loss matrix $i\Gamma$, wave function matching approach~\cite{song,mark} leads to a characteristic equation for eigenvalues of $H$ that results from the determinant of a 6$\times$6 matrix. It is, thus, of little analytical value to calculate the $\mathcal{PT}$-symmetric threshold $\gamma_{PT}(m)$ and instead, we obtain the $\mathcal{PT}$-symmetric phase diagram numerically. We restrict ourselves to the simplest case of a constant-tunneling Hamiltonian $H_0$ and an impurity potential matrix $i\Gamma=i\gamma_s\tau_z$ where $\tau_z$ is the $z$-Pauli matrix. In contrast to the previous cases, where the losses or gains for both modes occurred in the same waveguide, this non-Hermitian potential represents gain for one mode, $\sigma=+1$, and loss for the other, $\sigma=-1$, at site $m$. 

Figure~\ref{fig:phase} shows the numerically obtained results for the threshold $\gamma_{PT}(\mu)/t_s$ as a function of the fractional location $\mu=m/N$ of the first impurity for three different values of mode-mixing tunneling $t_d/t_s=\{0, 0.4, 0.7\}$. The left-hand panel shows the results for an even lattice with $N=40$. When $t_d=0$ (solid blue circles), the two degrees of freedom are uncoupled and $\mathcal{PT}$-symmetric phase diagram is identical to that for an open lattice with no internal degree of freedom~\cite{mark}. As $t_d/t_s$ increases (solid red squares and black stars), generically, we find that the critical $\gamma_{PT}(\mu)$ is non-monotonically suppressed for different values of impurity locations $\mu$. The right-hand panel shows corresponding results for an odd lattice with $N=41$. When $t_d=0$ (solid blue circles), the threshold impurity strength is given by $\gamma_{PT}/t_s=\sqrt{1+1/N}\approx 1.012$ when $m=1$~\cite{song}, and therefore, does not appear in the figure. Once again, when $t_d$ increases, the $\mathcal{PT}$-symmetric phase is (mostly) suppressed in a non-monotonic way. These results suggest that $\mathcal{PT}$-symmetry breaking in such systems shows a rich behavior that cannot be described with a simple analytical model. 
%--------------------------------------------------------------------------------%

\noindent{\it Discussion:} In this paper, we have introduced $\mathcal{PT}$-symmetric lattices with a local, two-state, quantum degree of freedom. By imposing invariance requirements on the Hermitian tunneling term, and $\mathcal{PT}$-symmetric potential term that represents spatially separated gain and loss impurities, we have shown that a broad class of such lattice systems can be expressed as the direct sum of two, uncoupled, $\mathcal{PT}$-symmetric systems. In such cases, we have predicted that $\mathcal{PT}$-symmetric threshold can be tuned by mode-mixing tunneling amplitude. Since the mapping is exact, all signatures of $\mathcal{PT}$-symmetry breaking, such as the ubiquitous, maximal chirality at $\mathcal{PT}$-symmetry breaking threshold~\cite{derekring}, the even-odd effect~\cite{evenodd}, tunable amplification~\cite{harsha}, etc. will be applicable in these cases as well. 

Since we have used the mode polarization as an example of the local degree of freedom, a microscopic calculation of the mode structure and the overlap between modes in adjacent waveguides is necessary to obtain typical tunneling matrix elements. Similarly a detailed study of the selection rules for different polarizations will be necessary to characterize the relative strengths of elements of the gain matrix $\gamma_s$ and $\gamma_d$. 
%--------------------------------------------------------------------------------%

\noindent{\it Acknowledgments:} This work was supported by the D.J. Angus-Scientech Educational Foundation (H.V.) and NSF Grant No. DMR-1054020 (Y.J.)

%---------------------------------------------------------------------------------%

%---------------------------------------------------------------------------------%
\end{document}